\documentclass[11pt]{iopart}

\usepackage[utf8]{inputenc}
\usepackage{graphicx,color}
\graphicspath{{figuras/}}
\usepackage{subfigure}
\usepackage{epstopdf}
\usepackage{cite} 
\usepackage{iopams}   
\usepackage[breaklinks=true]{hyperref}
\PassOptionsToPackage{linktocpage}{hyperref}

\begin{document}
\title{A simple and didactic method to calculate the fractal dimension - an interdisciplinary tool}

\author{P.V.S.Souza$^{1}$, R.L.Alves$^{1}$  e W.F.Balthazar$^{1}$ }

\address{$^1$Instituto Federal de Educação, Ciência e Tecnologia do Estado do Rio de Janeiro, Campus Volta Redonda, 27215-350 Volta Redonda - RJ, Brazil}

\ead{paulo.victor@ifrj.edu.br}
\vspace{10pt}
\begin{indented}
\item[]January 2018
\end{indented}

\begin{abstract}
Perfect fractals are mathematical objects that, because they are generated by recursive processes, have self-similarity and infinite complexity. In particular, they also have a fractional dimension. Although several proposals for the study of fractals at the basic level are present in the literature, only few proposals for the study of real fractals exist, which does not seem reasonable in our point of view considering the wealth of the theme and its potentiality as interdisciplinary theme in math education. For this reason, in this text, we present a simple and easily assimilable method to calculate the fractal dimension of any two-dimensional object. The method is divided into two steps. In the first step, students learn how is calculated the fractal dimension of any figure and they do this manually. Next part, which according to our best knowledge is new, with respect to the application in classroom, consists in the procedure of determining the fractal dimension using a smartphone and a computer with free software \textit {imageJ} installed. This proposal is easily understandable, considering the high school curriculum, and readily replicable, considering its easy adaptation to the most diverse school realities.
\end{abstract}

\maketitle
\ioptwocol
\section{Introduction}

Mathematics provides a set of abstract objects that are fundamental to the interpretation of the physical world. A very complex relationship between physics and mathematics exists, once mathematics is a organizer of physical knowledge \cite{doi:10.1179/030801811X13082311482609, 0031-9120-50-4-489}. In this sense, it is fundamental for the teaching of physics to think about approaching mathematical models that explain physical phenomena, collaborating for a better understanding of nature. In addition, we need realize that science has constantly innovated in the last decades, even so, a lot of current fundamental issues to the understanding of our world are not discussed in the high school classroom. For this reason, many papers propose that current topics be explored in physics classes \cite{0031-9120-44-5-011, 0031-9120-44-1-004, 0031-9120-48-2-238}. One of the topics wich may have important collaboration in the classroom, when it comes to mathematical modeling for understanding physical phenomena, is fractals. Because of its beauty, complexity and wide application, the study of these geometric forms constitute a powerful tool to obtain models that are closer to the phenomenology of nature \cite{mandelbrot1983fractal}.

There are several proposals that propose the discussion of the theme fractals in the area of mathematics teaching as well as physics teaching \cite{gomes87,sabin2009fractal,stavrou2008,knutson2003fractals, frame2002fractals,fraboni2008fractals, shriki2016fractals}. In spite of the wealth and utility of the theme, fractal geometry is practically absent from the  the curricula and from the textbooks of physics and mathematics of the high school. 

Considering the reader is familiar with the fundamental characteristics of the fractal objects\cite{mandelbrot1983fractal}, with the Hausdorff concept of dimension \cite{hausdorff18, falconer2004} and with the box-counting method to estimate the fractal dimension\cite{peitgen2006chaos,peitgen2012fractals, barnsley2014fractals}, in this paper, we present a simple and easily understandable method to calculate the fractal dimension of any two-dimensional figure \footnote{We provide a supplementary material with a concise discussion about the Hausdorff concept of dimension and  the box-counting method to estimate the fractal dimension in the following link \url{https://www.dropbox.com/s/riyxvuwo71lx1f9/SupplementaryMaterial.pdf?dl=0}.}. Basically, the method is divided into stages. Initially, students learn how is calculated the fractal dimension of any figure and they do this manually. The second part, which according to our best knowledge is new, with respect to the application in classroom, consists in the procedure of determining the fractal dimension using a smartphone and a computer with free software \textit {imageJ} installed. In this way, the student can calculate the fractal dimension of several figures and the concept of dimension can be discussed in the classroom.

We understand this procedure is ideal for high school and viable for the most diverse school realities. The method is easily assimilable and implementable, it can subsidize the teacher to work in an interdisciplinary way the theme fractals with many other subjects common to the high school curriculum. Moreover, we believe that the use of this didactic tool may provide students with the development of skills of wide application, such as the ability to use computational tools for image analysis. 

This text is organized as follows: initially, we present our experimental proposal for the fractal dimension study. This section is divided into two subsections where, first, we propose a manual activity and then a computational activity. Following, we present and discuss some results. Finally, we present some final remarks.

 
\section{Experimental activities and results} \label{results}

In this section, we present an experimental activity whose focus is the estimation of the fractal dimension of an object of any shape using the box count method. The activity is divided into two succeeding parts. In the first, the student must perform the counting of boxes manually to understand how the method is able to provide the Hausdorff dimension of an object of any shape. Then, with the help of free software, the same procedure is performed computationally, which allows us to apply it to the most diverse and complex objects and figures. For both activities we show experimental results for the Hausdorff dimension.

\subsection{The manual Box-Counting} 

We started with manual activity, which, due to the ease of the method, may be easily performed by high school students. This point is fundamental, once an intuitive and easy-to-understand method may be reproduced in any class. The central idea of this activity is provide to students the understanding about how the Hausdorff dimension is calculated by the box count method.

\begin{figure}[h]
	\centering
	\includegraphics[scale = 0.45]{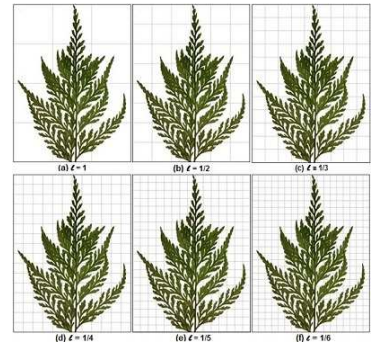}
	\caption{Example of manual application of the box count method.}
	\label{leaf}
\end{figure}

To illustrate, we present an example of estimating the fractal dimension of a leaf as shown in Figure \ref{leaf}. Initially, we printed several leaves with checkered grids, decreasing the size of each square and increasing the number of squares as many times as possible, our limit is determined by our visual ability. In the example under consideration, the grids were drawn with the aid of free software \textit {geogebra}, but could easily be made by hand. 

Considering $( {m})$ is the size of the square and ($ {N} $) is the number of squares occupied by the figure, we put the leaf over all grids, one at a time, and count the number of squares occupied by the image in each case. With this information we may construct the table below together with the dispersion diagram $\log {m}$ $\&$ $\log {N}$. The data in the graph may be adjusted by a linear function of type $y = ax + b$, where the angular coefficient ${a}$ is just the sought-dimension ${D}$. The fractional result for $ D = 1.69$ indicates that the figure takes up more space than a simple straight line and less space than a surface.

\begin{figure}[h]
	\centering
	\includegraphics[scale = 0.8]{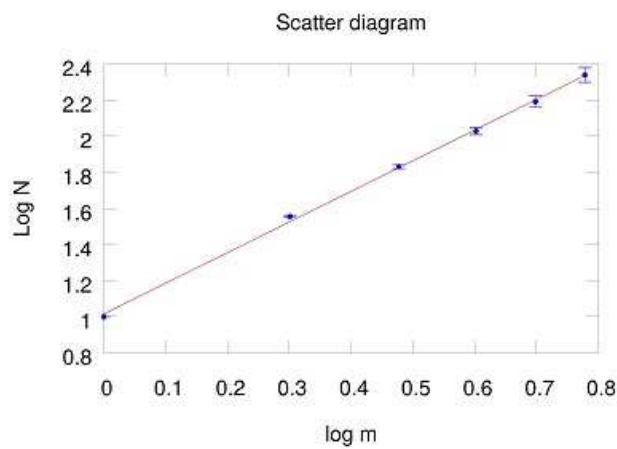}
	\caption{Scatter diagram $\log {m}$ x $\log{N}$. The experimental data are adjusted by the linear function $ f (x) = 1.69x + 1.01 $. The angular coefficient of the line corresponds to the dimension of the leaf. The error stems from the imprecision in the counting of some squares which, with the naked eye, it is not possible to say with certainty whether or not they are intercepted by the leaf.}
\end{figure}

On the one hand, as we may see, the manual counting method for calculating the Hausdorff dimension is easy to understand and offers very satisfactory results. On the other hand, manual counting has obvious limitations, which restricts its application to a more sophisticated problem. For instance, the accuracy of manual counting is limited to the brain's ability to compute to the naked eye. In addition, most everyday figures and objects are colored and/or presented in full view with different intensities of color. In this case, the application of box counting would require changes in the definition of fractal dimension itself, which is feasible for a specialist, but inconvenient or even impractical for the high school student \footnote {For example, one may calculate the fractal dimension of gray scale figures. However, the price is paid to modify the Haussdorf dimension definition previously presented in the text \cite{feder2013, Karperien2012}.}. In fact, while there are practical ways of circumventing the difficulties imposed by these nuances, they require extensive knowledge and, in general, are beyond the scope of high school courses. In response to this demand, we present in the next subsection a procedure of easy assimilation and replication that uses simple computational tools to estimate the fractal dimension of any figure in a way that is understandable for a high school student who has already performed the procedure manually.


\subsection{The Box-Counting method with computational resources}

To extend the application of the box-counting method - until now manual - to any image that one wants, we use computational resources. In fact, the use of computational resources and new technologies has subsidized the work of the teacher and provided a great advance in the process of teaching science learning, especially physics \cite{brown2009, bonato2017, klein2014}. In the case in question, we use FRACLAC - a plugin for free software, ImageJ, a Java-based and public domain image-processing program. The \textit {ImageJ} may be downloaded for free from the Internet at \url{https://imagej.nih.gov/ij/} and may run on any computer running a Java virtual machine 1.8 or higher. Currently, the program is available for platforms Windows, Mac OS X and Linux \cite{Karperien2012}.

The main question here concerns the function of the computational tool: FRACLAC calculates the fractal dimension of any figure from the box-counting process. This process, discussed earlier, may be easily understood by a high school student. In this way, the computational resource is not presented to the students as a kind of ``black box", rather, its use is made only to make a process which students are already fully familiar more complete and efficient. Thus, in this proposal, computational resources are not overestimated, but present themselves as tools that facilitate the treatment and study of the most diverse problems. However, most everyday figures and objects are colored and/or presented in full view with different color intensities. This is not a problem to use box counting with the software, but in this case the program works with gray levels and finds an average of intensity that serves as the threshold for binarizing the images. How then do you apply the box counting procedure in a way that is understandable to high school students? One way to do this is to subject the image to a \textit{binarization} process before counting boxes.  

The \textit{binarization} consists of a transformation of all pixels that make up a figure into white or black pixels, depending on the deviation of each pixel from the mean luminous intensity of the figure. This transformation may also be done very simply, with the help of the \textit{ImageJ} itself. With the software open, select the desired image in \textit{file} $ \rightarrow $ \textit{open}. The \textit{binarization} \hspace{2pt}process may be done by means of \textit{Process} $ \rightarrow $ \textit{Binary} $ \rightarrow $ \textit{Make Binary}.

\begin{figure}[h]
    \centering
            \includegraphics[scale = 0.25]{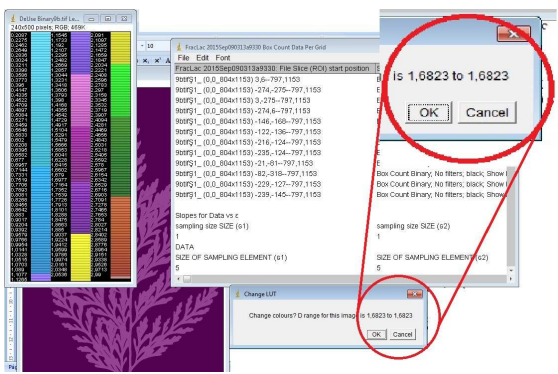}
            \label{fig:first_sub}
   \caption{The appearance of the \textit {imageJ} window once the process is finished. The value presented by the program as results of calculating the fractal dimension of the sheet using the box count method is highlighted.}
    \label{j1}
\end{figure}

After \textit{binarization}, the fractal dimension may be calculated. The FRACLAC plugin may be downloaded and installed following the guidelines presented in reference \cite{Karperien2012}. Once installed, the plugin may be accessed through the option \textit{Plugins $\rightarrow$ ``Fractal Analysis: FracLac"} \hspace{2pt} in the menu \textit{ImageJ}. Hence, we press \textit{BC} to set the box count. The user may then make changes or simply agree to default settings as done in this work. Then the \textit{Scan} button appears as an option. If pressed, the program calculates and displays the value of the fractal dimension, as shown in Figure \ref{j1}. We may verify that the $D$ range tells us the average of the DB values, the usual fractal counting dimension, on average, in relation to the number of scans that were made in the grid.

\begin{figure}[h]
	\centering
	\includegraphics[scale = 0.26]{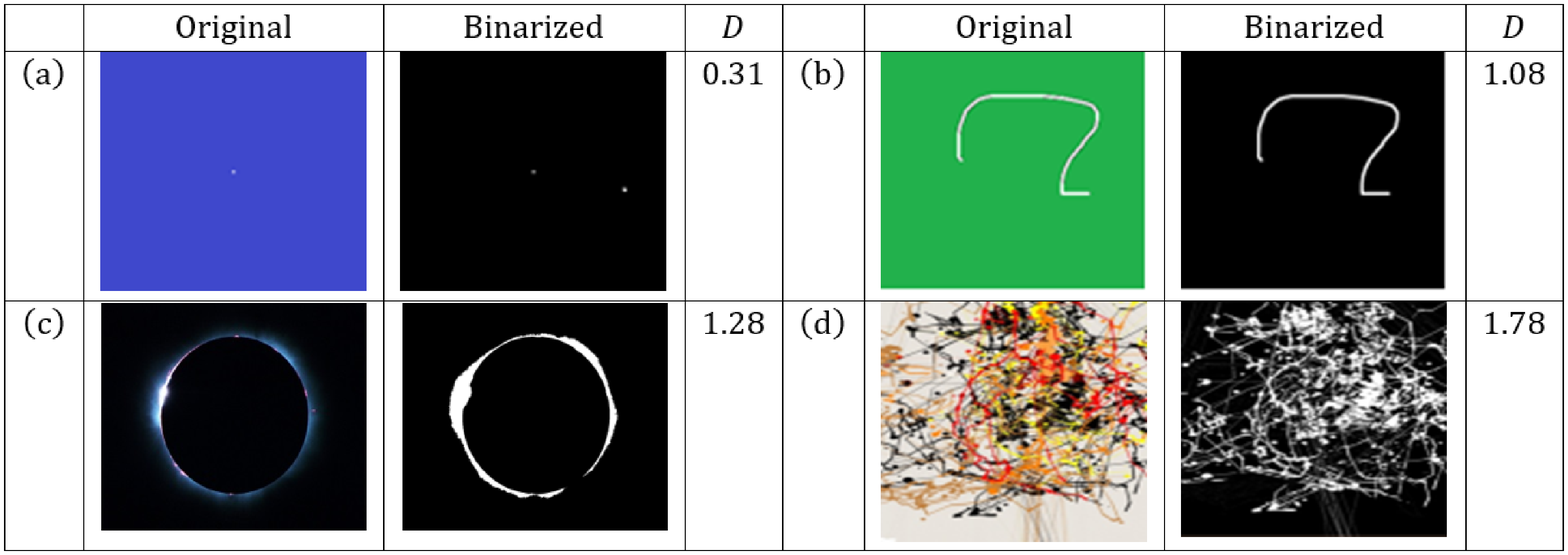}
	\caption{Original images, binarized images and the value of the fractal dimension $(D)$. In (a) we have an image produced by the authors in which a point appears in the middle of a plane; in (b), we have an image produced by authors in which a line appears in the middle of a plane; in (c), we have a photo of a solar eclipse \cite{eclipse}; finally, in (d) we have one of the most beautiful paintings by the american painter Jackson Pollock \cite{pollock}.}
	\label{Images}
\end{figure}

Using the computational method, we found the value $ 1.68$ to the dimension of the leaf, a value very close to the value found by the manual method, $ 1.69 $. This comparison is fundamental so that the student may have confidence in both methods presented. In order to illustrate the potentialities of the method, we present in Figure \ref{Images} some binarized examples with their respective dimensions calculated with the aid of \textit{ImageJ}.

Looking at the Figure \ref{Images} we notice that the dimension value grows as the difference between the black and white pixels in the image decreases. As already mentioned, the binarization process transforms all the pixels that make up the figure into black or white pixels, which, in general, ends up distinguishing the object (from which one wants to calculate the dimension) of the rest of the figure.

In fact, the program is not able to distinguish which part of the figure is the object and which part is the background, it only distinguishes the black and white pixels and supposes, by convention, that the object is composed by the least abundant number of them. Therefore, the program is of course not able to calculate the size of an entirely black or white figure. For example, in Figure \ref{Images} (a), the process calculates the size of the small central blur relative to the background landscape, we see that $ D $ is close to $ 0 $. In Figure \ref{Images} (b), we have the dimension of the string in relation to the background landscape, with a value close to $ 1 $. In Figure \ref {Images} (c) we have the dimension of the luminous band in relation to the darkness. Finally, in Figure \ref {Images} (d), we calculate the size of ink blots in relation to the screen. We may verify that the $ D $ dimension increases toward $ 2 $ when the figure, represented by white dots, is spreading along the black background.

It is important to mention that it would not be necessary to \textit{binarize} the figure before calculating its size through FRACLAC once the package itself already does this automatically. Our choice to \textit{binarize} the figures before calculating their dimension has a didactic objective, namely, to allow students to understand the process of \textit{binarization} that precedes the estimation of the fractal dimension.

We have thus noted that the method may be applied to the calculation of the fractal dimension of the most diverse figures, enabling and stimulating the approach of several themes present in the high school curriculum in an interdisciplinary perspective. We realize that the computational procedure is easily reproduced, because it demands only a few resources and may be accomplished in a short time, and understandable to the high school student, because it only automates and makes a more precise and efficient a manual procedure that the student already completely dominated.

\section{Final Remarks}

In this article, we present an experimental procedure that allows estimation of the fractal dimension of any figure. The proposal is carried out in two stages. In the first we have a manual process and in the second we use the software \textit {ImageJ}. We believe that both steps may be performed easily, allowing the student to understand what the fractal dimension is, as well as its importance for understanding nature's phenomena. In addition, the procedure requires little time and few computational resources, making it suitable to the most diverse school realities. 

It should be emphasized that the procedure may subsidize the teacher and provide tools for a more thorough and interesting investigation of several interdisciplinary themes. For example, one could use the procedure to discuss Olbers' paradox in astronomy classes, to estimate the size of coasts and watersheds in geography classes, to differentiate morphologically leaves in biology classes, to characterize roughness in optics classes, to investigate lightning in the atmosphere in electricity classes and even to discuss shade, light and perspective in photography and art classes. 

We believe that the activity proposed here
allows new topics in physics to be worked out in high schools. This means a great step not only for the innovation of the curriculum, but make possible teachers and students may transcend the concept of the dimension of Euclidean geometry, something that allows the development of a wider vision when we want to understand nature. In this sense, the authors' expectation is that the proposal allows students to understand and marvel at fractal geometry. 

This work is dedicated to Benoît Mandelbrot, who showed man the beauty and complexity of an entire Universe that can not be enclosed in three rigid dimensions.
\ack
The authors are indebted to Instituto Federal de Educação, Ciência e Tecnologia do Rio de Janeiro (IFRJ) for the incentive to research. This work was partially funded by Brazilian agency CNPQ (O Conselho Nacional de Desenvolvimento Científico e Tecnológico). We thanks A C F dos Santos for his useful comments.
\section{Bibliography}
\bibliographystyle{unsrt} 
\bibliography{ref}

\begin{thebibliography}{10}

\bibitem{doi:10.1179/030801811X13082311482609}
J~Lützen.
\newblock The physical origin of physically useful mathematics.
\newblock {\em Interdisciplinary {S}cience {R}eviews}, 36(3):229--243, 2011.

\bibitem{0031-9120-50-4-489}
C~Michelsen.
\newblock Mathematical modeling is also physics—interdisciplinary teaching
  between mathematics and physics in danish upper secondary education.
\newblock {\em Physics {E}ducation}, 50(4):489, 2015.

\bibitem{0031-9120-44-5-011}
W~M~S Santos, A~M Luiz, and C~R de~Carvalho.
\newblock A proposal to introduce a topic of contemporary physics into
  high-school teaching.
\newblock {\em Physics {E}ducation}, 44(5):511, 2009.

\bibitem{0031-9120-44-1-004}
S~Kapon, U~Ganiel, and B~Eylon.
\newblock Scientific argumentation in public physics lectures: bringing
  contemporary physics into high-school teaching.
\newblock {\em Physics {E}ducation}, 44(1):33, 2009.

\bibitem{0031-9120-48-2-238}
V~de~Souza, M~A Barros, E~C~Marques Filho, C~R Garbelotti, and H~A João.
\newblock Cosmic rays in the classroom.
\newblock {\em Physics {E}ducation}, 48(2):238, 2013.

\bibitem{mandelbrot1983fractal}
B~B Mandelbrot and R~Pignoni.
\newblock {\em The fractal geometry of nature}, volume 173.
\newblock WH freeman New York, 1983.

\bibitem{gomes87}
M~A~F Gomes.
\newblock Fractal geometry in crumpled paper balls.
\newblock {\em American {J}ournal of {P}hysics}, 55(7):649--650, 1987.

\bibitem{sabin2009fractal}
J~Sabin, M~Band{\'\i}n, G~Prieto, and F~Sarmiento.
\newblock Fractal aggregates in tennis ball systems.
\newblock {\em Physics {E}ducation}, 44(5):499, 2009.

\bibitem{stavrou2008}
D~Stavrou, R~Duit, and M~Komorek.
\newblock A teaching and learning sequence about the interplay of chance and
  determinism in nonlinear systems.
\newblock {\em Physics {E}ducation}, 43(4):417, 2008.

\bibitem{knutson2003fractals}
P~Knutson and E~D Dahlberg.
\newblock Fractals in the classroom.
\newblock {\em The Physics {T}eacher}, 41(7):387--389, 2003.

\bibitem{frame2002fractals}
M~Frame and B~Mandelbrot.
\newblock {\em Fractals, graphics, and mathematics education}.
\newblock Number~58. Cambridge University Press, 2002.

\bibitem{fraboni2008fractals}
M~Fraboni and T~Moller.
\newblock Fractals in the classroom.
\newblock {\em Mathematics {T}eacher}, 102(3):197--199, 2008.

\bibitem{shriki2016fractals}
A~Shriki and L~Nutov.
\newblock Fractals in the mathematics classroom: the case of infinite geometric
  series.
\newblock {\em Learning and {T}eaching {M}athematics}, 2016(20):38--42, 2016.

\bibitem{hausdorff18}
F~Hausdorff.
\newblock Dimension und {\"a}u{\ss}eres ma{\ss}.
\newblock {\em Mathematische Annalen}, 79(1):157--179, 1918.

\bibitem{falconer2004}
K~Falconer.
\newblock {\em Fractal geometry: mathematical foundations and applications}.
\newblock John Wiley \& Sons, 2004.

\bibitem{peitgen2006chaos}
H-O Peitgen, H~J{\"u}rgens, and D~Saupe.
\newblock {\em Chaos and fractals: new frontiers of science}.
\newblock Springer Science \& Business Media, 2006.

\bibitem{peitgen2012fractals}
H-O Peitgen, H~J{\"u}rgens, and D~Saupe.
\newblock {\em Fractals for the classroom: part two: complex systems and
  mandelbrot set}.
\newblock Springer Science \& Business Media, 2012.

\bibitem{barnsley2014fractals}
M~F Barnsley.
\newblock {\em Fractals everywhere}.
\newblock Academic press, 2014.

\bibitem{feder2013}
J~Feder.
\newblock {\em Fractals}.
\newblock Springer Science \& Business Media, 2013.

\bibitem{Karperien2012}
A~Karperien.
\newblock Fraclac for image{J}.
\newblock 2012.
\newblock Available at
  \url{https://imagej.nih.gov/ij/plugins/fraclac/FLHelp/Introduction.htm},
  acessed January 17, 2018.

\bibitem{brown2009}
D~Brown and A~J Cox.
\newblock Innovative uses of video analysis.
\newblock {\em The Physics Teacher}, 47(3):145--150, 2009.

\bibitem{bonato2017}
J~Bonato, L~M Gratton, P~Onorato, and S~Oss.
\newblock Using high speed smartphone cameras and video analysis techniques to
  teach mechanical wave physics.
\newblock {\em Physics Education}, 52(4):045017, 2017.

\bibitem{klein2014}
P~Klein, S~Gröber, J~Kuhn, and A~Müller.
\newblock Video analysis of projectile motion using tablet computers as
  experimental tools.
\newblock {\em Physics Education}, 49(1):37, 2014.

\bibitem{eclipse}
The Landscape~Photography Podcast.
\newblock Eclipse, 2018.
\newblock Available at
  \url{https://www.landscapephotographypodcast.com/podcast/2017/8/6/landscape-photography-podcast-ep-8},
  acessed January 17, 2018.

\bibitem{pollock}
Democrart.
\newblock Jackson pollock's painting, 2018.
\newblock Available at
  \url{http://www.democrart.com.br/aboutart/artista/jackson-pollock/}, acessed
  January 17, 2018.

\end{thebibliography}
\end{document}